\documentclass[twocolumn]{autart}
\usepackage[utf8]{inputenc}
\usepackage{graphicx}
\usepackage{caption}
\usepackage{tabularx,booktabs}

\usepackage{multicol}
\usepackage{amsmath,amssymb}
\usepackage{amsfonts}
\usepackage{bbm}
\usepackage{textcomp}
\usepackage{psfrag}
\usepackage{multimedia}
\usepackage{fancybox}

\newtheorem{theorem}{Theorem}

\newtheorem{Lemma}{Lemma}
\newtheorem{Definition}{Definition}
\newtheorem{Remark}{Remark}

\newtheorem{Assumption}{Assumption}

\usepackage[usenames,dvipsnames]{xcolor}

\renewcommand\refname{References and Notes}

\newcommand {\matr}[2]{\left[\begin{array}{#1}#2\end{array}\right]}

\newcommand{\vect}[1]{{\ensuremath{\boldsymbol{{#1}}}}}

\newcounter{lastnote}

\begin{document} 

\begin{frontmatter}
	
	\title{A New Dissipativity Condition for Asymptotic Stability of Discounted Economic MPC\thanksref{footnoteinfo}}

	\thanks[footnoteinfo]{This paper was not presented at any IFAC 
		meeting. Corresponding author M. Zanon. This paper was partially supported by the Italian Ministry of University and Research under the PRIN'17 project ``Data-driven learning of constrained control systems" , contract no. 2017J89ARP; by ARTES 4.0 Advanced Robotics and enabling digital Technologies \& Systems 4.0, CUP: B81J18000090008; and by the Norwegian Research Council project ``Safe Reinforcement-Learning using MPC" (SARLEM).}
	
	\author[Mario]{Mario Zanon}\ead{mario.zanon@imtlucca.it},
	\author[Seb]{S\'ebastien Gros}, 
	
	\address[Mario]{IMT School for Advanced Studies Lucca, Piazza San Francesco 19, 55100, Lucca, Italy}
	\address[Seb]{NTNU, Gløshaugen, Trondheim, Norway}
	
	
	\begin{abstract}
		Economic Model Predictive Control has recently gained popularity due to its ability to directly optimize a given performance criterion, while enforcing constraint satisfaction for nonlinear systems. Recent research has developed both numerical algorithms and stability analysis for the undiscounted case. The introduction of a discount factor in the cost, however, can be desirable in some cases of interest, e.g., economics, stochastically terminating processes, Markov decision processes, etc. Unfortunately, the stability theory in this case is still not fully developed. In this paper we propose a new dissipativity condition to prove asymptotic stability in the infinite horizon case and we connect our results with existing ones in the literature on discounted economic optimal control. Numerical examples are provided to illustrate the theoretical results.
	\end{abstract}
	
\end{frontmatter}

\section{Introduction}

Model Predictive Control (MPC) has become popular thanks to its ability to control nonlinear systems while explicitly imposing constraint satisfaction and optimizing a given objective~\cite{Grune2011,Rawlings2017}. However, the ability to optimize a given objective is only partially exploited in \emph{tracking MPC}, where the cost penalizes the distance from a given (possibly optimal) reference. On the contrary, in so-called \emph{economic MPC} a given performance criterion is directly optimized, with the aim to achieve optimal performance not only at the reference, but also during transients.

The main drawback of economic MPC is the increased difficulty in guaranteeing asymptotic stability, compared to the tracking MPC case. This difficulty stems from the fact that the stage cost is generic and not necessarily positive-definite with respect to a given steady-state. Recent research has developed a sound stability theory for undiscounted economic MPC based on the concept of \emph{strict dissipativity}~\cite{Amrit2011a,Diehl2011,Faulwasser2018a,Faulwasser2018,Grune2013a,Gruene2014,Mueller2015a,Zanon2013d,Zanon2014d,Zanon2016b,Zanon2017a,Zanon2017e,Zanon2018a}.

While MPC is commonly formulated in an undiscounted setting, in a number of cases it makes sense to introduce a discount factor. The discount factor can carry several meanings, e.g., as a model of the interest rate in case the problem consists in maximizing a given capital; or as a model of processes having an uncertain lifetime~\cite{Puterman1994}. Additionally, a discount factor is often the preferred approach to formulate well-posed Markov Decision Processes (MDPs), which offer one of the most generic formulation for optimal control. In particular, MDPs are at the core of Reinforcement Learning techniques~\cite{Sutton2018}, which are most developed for the discounted case, while the undiscounted setting has received less attention.

Unfortunately, the stability theory for discounted MPC formulations is more involved than the one for the undiscounted setting. As proven in~\cite{Postoyan2014}, even the case of a positive-definite stage cost over an infinite horizon poses additional difficulties and is nontrivial to analyze. The case of an economic cost over an infinite horizon has been analyzed in, e.g.,~\cite{Gaitsgory2018,Gruene2016,Gruene2020}. In~\cite{Gruene2020a} the authors propose a local stability analysis which makes it possible to characterize cases in which the optimal steady-state depends on both the initial state and the discount factor.  Unfortunately, a tight condition for stability is still not available in the literature, as we will discuss more in detail in Section~\ref{sec:strong_discounted_strict_dissipativity}. 

In this paper, we propose a new strict dissipativity concept for discounted infinite-horizon MPC which is stronger than existing strict dissipativity concepts but allows us to prove asymptotic stability without additional conditions. 
Moreover, we provide a characterization of the optimal steady-state to which the system is stabilized. Finally, we connect the discounted case to the undiscounted one by proving that the discount factor can be eliminated, provided that the cost is suitably modified. Our assumptions allow us to provide a different insight than the one provided in~\cite{Gaitsgory2018}, though we will show that the two sets of assumption share many common points. A thorough investigation of the necessity of our assumptions for asymptotic stability will be the subject of future research.

This paper is organized as follows. We introduce the problem and briefly recall the main strict dissipativity definitions in Section~\ref{sec:preliminaries}. We define the new strict dissipativity concept, prove asymptotic stability and characterize the optimal steady-state in Section~\ref{sec:strong_discounted_strict_dissipativity}, where we also compare our assumptions to those of~\cite{Gaitsgory2018}, which studies the same setting. We provide some examples in Section~\ref{sec:examples} and conclude with Section~\ref{sec:conclusions}.

\section{Preliminaries}
\label{sec:preliminaries}

We consider nonlinear discrete-time systems with dynamics and stage cost given respectively by
\begin{align}
	\vect{x}_+ = \vect{f} \left ( \vect{x},\vect{u} \right ), &&\text{and} && L(\vect{x},\vect{u}),
\end{align}
where $\vect{x}\in\mathbb{R}^{n_\vect{x}}$, $\vect{u}\in\mathbb{R}^{n_\vect{u}}$ denote the states and inputs respectively. Since we are interested in a discounted optimal control setting, we further introduce the discount factor $\gamma\in]0,1[$. Furthermore, the system is subject to the state and input constraints
\begin{align*}
	\left (\vect{x},\vect{u}\right ) \in \mathbb{Z} := \left \{\, \left (\vect{x},\vect{u}\right ) \,|\, \vect{h}\left (\vect{x},\vect{u}\right ) \leq 0 \,\right \}.
\end{align*}
For ease of notation and without loss of generality, we will assume that $L(\vect{x},\vect{u})=\infty$ for all $\left (\vect{x},\vect{u}\right ) \notin \mathbb{Z}$.

For a given policy $\vect{\pi}(\vect{x})$, we will denote closed-loop trajectories as
\begin{align}
	\vect{x}_{k+1}^\vect{\pi} = \vect{f} \left ( \vect{x}_{k}^\vect{\pi},\vect{\pi}\left (\vect{x}_{k}^\vect{\pi}\right ) \right ), && \vect{x}_0^\vect{\pi} = \vect{x}_0,
\end{align}
where we omit the dependence on $\vect{x}_0$, since the initial state is independent of the policy and no confusion can arise. We define the sets of admissible initial conditions and policies as
\begin{align*}
	\mathbb{X}_0 &:= \left \{\, \vect{x}_0 \,|\,  \exists \,\vect{\pi} \ \mathrm{s.t.} \ \vect{h}\left (\vect{x}_k^{\vect{\pi}},\vect{\pi}\left (\vect{x}_k^{\vect{\pi}}\right )\right ) \leq 0, \forall \, k\geq0  \,\right \}, \\
	\Pi &:= \left \{ \, \vect{\pi} \, | \,  \vect{h}\left (\vect{x}_k^{\vect{\pi}},\vect{\pi}\left (\vect{x}_k^{\vect{\pi}}\right )\right ) \leq 0, \forall \, \vect{x}_0 \in \mathbb{X}_0, \, k \geq0 \, \right \}.
\end{align*}

\begin{Assumption}
	\label{ass:regularity}
	Sets $\mathbb{Z}$ and $\mathbb{X}_0$ are compact. Moreover, $|L(\vect{x},\vect{u})|<\infty, \ \forall \, (\vect{x},\vect{u})\in\mathbb{Z}$. 
\end{Assumption}
The assumption on boundedness of $L$ is without loss of generality, since one can exclude all points for which it becomes unbounded by suitably defining $\vect{h}$ and, consequently, $\mathbb{Z}$.
The role of the compactness assumption on $\mathbb{Z}$ and $\mathbb{X}_0$ is to guarantee that the functions we will use when defining the cost remain bounded on the system trajectories. While this assumption can be relaxed, we adopt it here in order to avoid additional technicalities. 

For a given discount factor $\gamma \in ]0,1[$, we define the optimal value function of the MPC problem as
\begin{align}
	V_\star^\gamma \left (\vect{x}_0\right ) := \min_{\vect{\pi}\in \Pi} \ \sum_{k=0}^\infty \gamma^k L\left (\vect{x}_k^\vect{\pi},\vect{\pi}\left (\vect{x}_k^\vect{\pi}\right )\right ), \label{eq:ValueFunction}
\end{align}
and denote any optimal policy as $\vect{\pi}_\star^\gamma$. 
We recall that the optimal value function \eqref{eq:ValueFunction} satisfies the Bellman equation
\begin{align}
	\label{eq:bellman}
	V_\star^\gamma \left (\vect{x}\right ) = \min_{\vect{\pi}\in \Pi} \ L(\vect{x},\vect{\pi}(\vect{x})) + \gamma 	V_\star^\gamma \left (\vect{f} \left ( \vect{x},\vect{\pi}(\vect{x})\right )\right ).
\end{align}
In this paper we will compare several infinite-horizon MPC formulations of the form~\eqref{eq:ValueFunction}, which differ by the used discount factor or stage cost. With slight abuse of notation, though $\vect{\pi}_\star^\gamma$ refers to one of the possibly many policies which are optimal for~\eqref{eq:ValueFunction}, we will write, e.g.,  $\vect{\pi}=\vect{\pi}_\star^\gamma$ to denote that a given policy $\vect{\pi}$ belongs to the set of optimal policies of~\eqref{eq:ValueFunction}. Moreover, throughout the paper we will use the following definition.
\begin{Definition}[MPC Equivalence]
	Two MPC formulations are \emph{equivalent} if their value functions coincide and any policy which is optimal for one MPC formulation is also optimal for the other MPC formulation.
\end{Definition}

In the following, we will make use of the comparison functions defined by the function classes $\mathcal{K}$, $\mathcal{K}_\infty$, $\mathcal{L}$, $\mathcal{KL}$. We define $\mathbb{R}_{\geq0}:=\{ \ x \in\mathbb{R} \ | \ x\geq0 \ \}$. Function $\alpha:\mathbb{R}_{\geq0}\to\mathbb{R}_{\geq0}$ satisfies $\alpha\in\mathcal{K}$ if it is continuous, zero at zero and strictly increasing. If additionally $\alpha\in\mathcal{K}$ is radially unbounded, then $\alpha\in\mathcal{K}_\infty$. Function $\delta:\mathbb{R}_{\geq0}\to\mathbb{R}_{\geq0}$ satisfies $\delta\in\mathcal{L}$ if it is continuous and strictly decreasing with $\lim_{t\to\infty} \delta(t)=0$. Finally, $\beta:\mathbb{R}_{\geq0}\times\mathbb{R}_{\geq0}\to\mathbb{R}_{\geq0}$ satisfies $\beta\in\mathcal{KL}$ if it is continuous, $\beta(\cdot,t)\in\mathcal{K}$, and $\beta(x,\cdot)\in\mathcal{L}$.

In this paper, we aim at characterizing the tuples $\left(\vect f, L , \gamma\right)$ that yield optimal policies $\vect{\pi}_\star^\gamma$ achieving asymptotic stability to a given steady state. To that end, we need the following definition. 
\begin{Definition}
	The steady state 
	\begin{align}
		\vect{x}^\mathrm{s} = \vect{f} \left ( \vect{x}^\mathrm{s},\vect{u}^\mathrm{s}\right )
	\end{align}
	is asymptotically stable if there exists a function $\beta\in\mathcal{KL}$ such that all closed-loop trajectories satisfy
	\begin{align}
		\left \|\vect{x}_k^{\vect{\pi}_\star^\gamma} - \vect{x}^\mathrm{s} \right \| \leq \beta\left ( \left \|\vect{x}_0 - \vect{x}^\mathrm{s}\right \|, k \right ).
	\end{align}
\end{Definition}
Given a steady state $\left ( \vect{x}^\mathrm{s},\vect{u}^\mathrm{s}\right )$ candidate for proving asymptotic stability, without loss of generality and for simplicity of notation we will assume throughout the paper that $L(\vect{x}^\mathrm{s},\vect{u}^\mathrm{s})=0$. 

Since our analysis will be constructed based on a newly defined dissipativity concept, we summarize next existing dissipativity definitions used to prove stability in the MPC context.

\subsection{A Brief Summary on Dissipativity and Stability}

In order to study the stability properties of problems for which $L(\vect{x},\vect{u}) \ngeq \alpha(\| \vect{x}-\vect{x}^\mathrm{s} \|)$ for some $\vect{x}^\mathrm{s}$, $\alpha\in\mathcal{K}$, the following concept of \textit{strict dissipativity} has been introduced in~\cite{Amrit2011a,Diehl2011} for the case of $\gamma=1$:
\begin{align}
	\label{eq:strict_diss}
	\exists \, \lambda :  \  L(\vect{x},\vect{u}) + \lambda(\vect{x}) - \lambda(\vect{f}(\vect{x},\vect{u})) \geq\rho(\|\vect{x}-\vect{x}^\mathrm{s}\|),
\end{align}
for $\rho\in\mathcal{K}$.
Under this condition, if $\lambda$ is bounded, asymptotic stability has been proven in~\cite{Amrit2011a}. Necessity of strict dissipativity has been proven in~\cite{Mueller2015a} under an additional controllability assumption, and provided that $(\vect{x}^\mathrm{s},\vect{u}^\mathrm{s})\in\mathrm{int}(\mathbb{Z})$. 
This dissipativity concept has been extended in several directions, including considering the periodic case~\cite{Mueller2016,Zanon2017e}, and approximate economic MPC schemes~\cite{Zanon2014d,Zanon2016b,Zanon2017a}. 
Additionally, some results require strictness also in the inputs, i.e., $\rho(\|\vect{x}-\vect{x}^\mathrm{s}\|)$ is replaced by $\rho(\|\vect{x}-\vect{x}^\mathrm{s}, \ \vect{u}-\vect{u}^\mathrm{s} \|)$. For more details on the topic we refer to the excellent survey~\cite{Faulwasser2018a} and references therein.

In the context of discounted optimal control, the conditions for stability are harder to derive. Relevant work includes~\cite{Postoyan2014}, where the stage cost is assumed to be positive-definite, and~\cite{Gaitsgory2018,Gruene2016,Gruene2020}, which rely on the following \emph{discounted strict dissipativity} concept:
\begin{align}
	\label{eq:discointed_strict_diss}
	\exists \, \lambda : \  L(\vect{x},\vect{u}) + \lambda(\vect{x}) - \gamma \lambda(\vect{f}(\vect{x},\vect{u})) \geq\rho(\|\vect{x}-\vect{x}^\mathrm{s}\|).
\end{align}
In this case, the discounted strict dissipativity condition is not sufficient to prove stability. We will discuss further in Section~\ref{sec:stability_conditions} the required additional assumptions, and observe that in~\cite{Gaitsgory2018} practical stability is proven, rather than asymptotic stability.

We propose next a new dissipativity concept which can be interpreted as the joint conditions~\eqref{eq:strict_diss}-\eqref{eq:discointed_strict_diss}, where the stage cost in the first condition needs to be modified. Since the term discounted strict dissipativity is in use to define condition~\eqref{eq:discointed_strict_diss}, we will refer to our new condition as \emph{Strong Discounted Strict Dissipativity} (SDSD), since it implies~\eqref{eq:discointed_strict_diss} but the converse is in general not true.

\section{Strong Discounted Strict Dissipativity}
\label{sec:strong_discounted_strict_dissipativity}

In this section, we consider the stabilization to an arbitrary steady-state $(\vect{x}_\mathrm{s},\vect{u}_\mathrm{s})$. To that end, we will first define SDSD; then, assuming that it holds for $(\vect{x}_\mathrm{s},\vect{u}_\mathrm{s})$, we will prove that it entails asymptotic stability of the closed-loop system to $(\vect{x}_\mathrm{s},\vect{u}_\mathrm{s})$. Afterwards, we will further specify the set of steady-states to which the system can be stabilized, and describe that set in terms of a steady-state optimality criterion. To that end, we will further establish an equivalence between the discounted and undiscounted case. Finally, we will compare our assumption to an alternative one used in the literature.

Strong Discounted Strict Dissipativity holds if the tuple $\left(\vect f, L , \gamma\right)$ satisfies the following assumption.
\begin{Assumption}[SDSD]
	\label{ass:disc_str_diss}
	There exist a function $\lambda(\vect{x})$, continuous at $\vect{x}^\mathrm{s}$, bounded for bounded $\vect{x}$, satisfying $\lambda(\vect{x}^\mathrm{s})=0$, and a function $\rho\in\mathcal{K}$ such that:
	\begin{subequations}
		\label{eq:discounted_dissipativity}
		\begin{align}
		(i) \qquad &L(\vect{x},\vect{u}) + \lambda(\vect{x}) - \gamma \lambda(\vect{f}(\vect{x},\vect{u})) \geq \rho(\|\vect{x}-\vect{x}^\mathrm{s}\|), \label{eq:discounted_dissipativity1}\\
		(ii)\qquad &L(\vect{x},\vect{u}) + \lambda(\vect{x}) - \lambda(\vect{f}(\vect{x},\vect{u})) \nonumber \\
		&\hspace{3.25em}+ (\gamma -1)V_\star^\gamma (\vect{f}(\vect{x}, \vect{u})) \geq \rho(\|\vect{x}-\vect{x}^\mathrm{s}\|), \label{eq:discounted_dissipativity2}
		\end{align}
	\end{subequations}
	hold for all $(\vect x,\vect u)\in\mathbb{Z}$.
\end{Assumption}
In the limit $\gamma\to1$, the two conditions coincide and we recover the standard definition of strict dissipativity for the undiscounted case~\eqref{eq:strict_diss}. Furthermore, condition~\eqref{eq:discounted_dissipativity1} coincides with discounted strict dissipativity~\eqref{eq:discointed_strict_diss}. 

In the more general case in which the states can also take unbounded values, additional assumptions need to be introduced to guarantee boundedness of $\lambda(\vect{x})$, and the stronger requirement $\rho\in\mathcal{K}_\infty$ is needed. Since the remainder of our analysis remains valid in that case, we avoid these technicalities for the sake of simplicity. 

Finally, we introduce a standard assumption which can be interpreted as a weak controllability assumption for cost functions which are sufficiently regular~\cite{Rawlings2017}.
\begin{Assumption}
	\label{ass:controllability}	
	The value function $V_\star^\gamma (\vect{x})$
	is continuous at $\vect{x}^\mathrm{s}$, and bounded on $\mathbb{X}_0$. 
\end{Assumption}
We are now ready to deliver our main results.

\subsection{Asymptotic Stability}

In order to prove asymptotic stability, we will first define a modified stage cost and recall an existing result which will be exploited in the proof.

We define the following modified cost\footnote{in this case we do not adopt the nomenclature ``rotated cost'' used in the undiscounted case, which we reserve for a different definition of stage cost used later.}, together with the associated optimal policy and value function:
%\begin{subequations}
	\begin{align}
		\hspace{-3pt}\hat L^\gamma(\vect{x},\vect{u}) &:= L(\vect{x},\vect{u}) +  \lambda(\vect{x}) - \gamma \lambda(\vect{f}(\vect{x},\vect{u})), \label{eq:Lhat}\\
		\vect{\hat \pi}_\star^\gamma \left (\vect{x}_0\right ) &:= \arg\min_{\vect{\pi}\in\Pi} \ \sum_{k=0}^\infty \gamma^k \hat L^\gamma \left (\vect{x}_k^\vect{\pi},\vect{\pi}\left (\vect{x}_k^\vect{\pi}\right )\right ) 
		. \label{eq:pihat} \\
		\hat V_{\vect{\pi}}^\gamma \left (\vect{x}_0\right ) &:= \sum_{k=0}^\infty \gamma^k \hat L^\gamma \left (\vect{x}_k^\vect{\pi},\vect{\pi}\left (\vect{x}_k^\vect{\pi}\right )\right ), \ \ \hat V_\star^\gamma (\vect{x}) := \hat V_{\vect{\hat \pi}_\star^\gamma}^\gamma (\vect{x}). \label{eq:Vhat}
	\end{align}
%\end{subequations}

Similarly to the undiscounted case~\cite{Amrit2011a,Diehl2011}, the cost modification that yields $\hat L^\gamma$ does not affect the primal solution, as stated in the next theorem.
\begin{theorem}[\cite{Gruene2016},~\cite{Gruene2020}]
	Suppose that $\lambda$ is bounded. Then, for all $\vect{x}_0\in\mathbb{X}_0$, we have $\vect{\hat \pi}_\star^\gamma (\vect{x})=\vect{\pi}_\star^\gamma (\vect{x})$, and
	\begin{align}
		\label{eq:V_Vhat}
		\hat V_\star^\gamma(\vect{x})=V_\star^\gamma(\vect{x}) + \lambda(\vect{x}).
	\end{align}
\end{theorem}

We are now ready to prove our main result.
\begin{theorem}[Asymptotic Stability]
	\label{thm:asymptotic_stability}
	Suppose that Assumptions~\ref{ass:regularity}, \ref{ass:disc_str_diss}, \ref{ass:controllability} hold. 
	Then, for all $\vect{x}_0\in\mathbb{X}_0$, 
	$\hat V_\star^\gamma$ is a Lyapunov function and the system in closed-loop with policy $\vect{\hat \pi}_\star^\gamma (\vect{x})=\vect{\pi}_\star^\gamma (\vect{x})$ is asymptotically stabilized to the steady state $\vect{x}^\mathrm{s}$.
\end{theorem}
\begin{pf}
	Using \eqref{eq:discounted_dissipativity1}, we obtain
		\begin{align}
	&\hat L^\gamma (\vect{x},\vect{u}) \geq \rho(\|\vect{x}-\vect{x}^\mathrm{s}\|), \label{eq:discounted_dissipativity_hat1}
	\end{align}
	such that $\hat V_\star^\gamma(\vect{x})\geq \rho(\|\vect{x}-\vect{x}^\mathrm{s}\|)$. 

	Because by assumption $V_\star^\gamma$ and $\lambda$ are bounded, zero and continuous at $\vect{x}^\mathrm{s}$, then also $\hat V_\star^\gamma$ is bounded, zero and continuous at $\vect{x}^\mathrm{s}$. Therefore, it can be upper-bounded by a $\mathcal{K}$ function $\alpha$ such that~\cite[Proposition B.25]{Rawlings2017}:
	\begin{align}
		\hat V_\star^\gamma(\vect{x}) \leq \alpha(\| \vect{x}-\vect{x}^\mathrm{s} \|).
	\end{align}
Using~\eqref{eq:Lhat} and~\eqref{eq:V_Vhat}, Condition~\eqref{eq:discounted_dissipativity2} equivalently reads as
	\begin{align}
		&\hat L^\gamma (\vect{x},\vect{u}) +(\gamma-1) \left (V_\star^\gamma (\vect{f}(\vect{x}, \vect{u}))+ \lambda (\vect{f}(\vect{x}, \vect{u})) \right ) \nonumber \\
		&\hspace{2em}\overset{\eqref{eq:V_Vhat}}{=}\hat L^\gamma (\vect{x},\vect{u}) +(\gamma-1) \hat V_\star^\gamma (\vect{f}(\vect{x}, \vect{u})) \geq \rho(\|\vect{x}-\vect{x}^\mathrm{s}\|), \label{eq:discounted_dissipativity_hat2}
	\end{align}
	which entails that $\hat V_\star^\gamma$ satisfies the decrease condition
	\begin{align}
	&\hat V_\star^\gamma\left  (\vect{f}\left (\vect{x}, \vect{\pi}_\star^\gamma\left (\vect{x}\right )\right )\right ) - \hat V_\star^\gamma (\vect{x}) \nonumber \\
	&\hspace{2em}= -\left (\hat L^\gamma \left (\vect{x},\vect{\pi}_\star^\gamma\left (\vect{x}\right )\right ) +(\gamma-1) \hat V_\star^\gamma \left (\vect{f}\left (\vect{x}, \vect{\pi}_\star^\gamma\left (\vect{x}\right )\right )\right )  \right )\nonumber \\
	&\hspace{2em}\leq - \rho\left (\left \|\vect{x}-\vect{x}^\mathrm{s}\right \|\right ). \label{eq:lyap_decrease}
	\end{align}
	$\hfill\qed$
\end{pf}

After having proved asymptotic stability, we now turn to the characterization of the set of optimal steady states, to which the system is stabilized if SDSD holds. 
To that end, we will first establish an equivalence between the discounted MPC and a suitably defined undiscounted MPC.

\subsection{Equivalence with the Undiscounted Case}

In this section, we formulate an undiscounted formulation which yields the same policy and value function as the discounted one.
To that end, we define a new modified stage cost, which differs from~\eqref{eq:Lhat} and, as we will discuss later, is related to condition~\eqref{eq:discounted_dissipativity2}: 
\begin{align}
	\label{eq:undiscounted_equivalent_stage_cost}
	\tilde L^\gamma(\vect{x},\vect{u}) &:= L(\vect{x},\vect{u}) + (\gamma -1)V_\star^\gamma (\vect{f}(\vect{x}, \vect{u})),
\end{align}
with corresponding undiscounted value function and optimal policy 
\begin{align}
	\vect{\tilde \pi}_\star^\gamma \left (\vect{x}_0\right ) &:= \arg\min_{\vect{\pi}\in\Pi} \ \sum_{k=0}^\infty \tilde L^\gamma \left (\vect{x}_k^\vect{\pi},\vect{\pi}\left (\vect{x}_k^\vect{\pi}\right )\right ) 
	, \label{eq:pitilde} \\
	\tilde V_{\vect{\pi}}^\gamma \left (\vect{x}_0\right ) &:= \sum_{k=0}^\infty \tilde L^\gamma\left (\vect{x}_k^\vect{\pi},\vect{\pi}\left (\vect{x}_k^\vect{\pi}\right )\right ), \ \  \tilde V_\star^\gamma (\vect{x}) := \tilde V_{\vect{\tilde \pi}_\star^\gamma}^\gamma (\vect{x}). \label{eq:Vtilde}
\end{align}

\begin{theorem}
	\label{thm:undiscounted_equivalence}
	Suppose that Assumptions~\ref{ass:regularity},~\ref{ass:disc_str_diss},~\ref{ass:controllability} hold. Then, for all $\vect{x}_0\in\mathbb{X}_0$,
	\begin{align}
		\vect{\tilde \pi}_\star^\gamma(\vect{x}) = \vect{ \pi}_\star^\gamma(\vect{x}), &&  \tilde V_\star^\gamma(\vect{x})= V_\star^\gamma(\vect{x}).
	\end{align}
\end{theorem}
\begin{pf}
	We use~\eqref{eq:undiscounted_equivalent_stage_cost} and~\eqref{eq:bellman} to write
	\begin{align*}
		\tilde L^\gamma \left (\vect{x},\vect{\pi}_\star^\gamma(\vect{x})\right ) &= V_\star^\gamma (\vect{x}) -  V_\star^\gamma \left (\vect{f}\left (\vect{x},\vect{\pi}_\star^\gamma(\vect{x})\right )\right ),
	\end{align*}
	which we use in~\eqref{eq:Vtilde} to obtain, for all $\vect x_0\in\mathbb X_0$,
	\begin{align}
		\label{eq:Vhat_V}
		\tilde V_{\vect{\pi}_\star^\gamma}^\gamma (\vect{x}_0) &= \sum_{k=0}^\infty V_\star^\gamma \left (\vect{x}_k^{\vect{\pi}_\star^\gamma}\right ) -  V_\star^\gamma \left (\vect{f}\left (\vect{x}_k^{\vect{\pi}_\star^\gamma},\vect{\pi}_\star^\gamma\left (\vect{x}_k^{\vect{\pi}_\star^\gamma}\right )\right )\right ) \nonumber \\
		&= V_\star^\gamma \left (\vect{x}_0\right ),
	\end{align}
	where we simplified the terms in the telescopic sum, which are all finite since $\vect x_0\in\mathbb X_0$ and Assumptions~\ref{ass:regularity}-\ref{ass:disc_str_diss} hold, and we exploited Theorem~\ref{thm:asymptotic_stability}, $\lambda\left(\vect x^\mathrm{s}\right)=0$ and~\eqref{eq:V_Vhat} to establish $\lim_{k\to\infty}  V_\star^\gamma\left (\vect{x}_k^{\vect{\pi}_\star^\gamma}\right )=0$. 
	Then, we apply the Dynamic Programming recursion, and exploit~\eqref{eq:undiscounted_equivalent_stage_cost},~\eqref{eq:Vhat_V} to derive
	\begin{align*}
		&\arg\min_{\vect{\pi}\in\Pi} \ \tilde L^\gamma (\vect{x},\vect{\pi}(\vect{x})) + \tilde V_{\vect{\pi}_\star^\gamma}^\gamma (\vect{f}(\vect{x},\vect{\pi}(\vect{x}))) \\
		&\hspace{1em} =  \arg\min_{\vect{\pi}\in\Pi} \ L(\vect{x},\vect{\pi}(\vect{x})) + \gamma V_\star^\gamma (\vect{f}(\vect{x},\vect{\pi}(\vect{x}))) 
		=  \vect{\pi}_\star^\gamma(\vect{x}),
	\end{align*}
	which proves that $\vect{\pi}_\star^\gamma$ is optimal for cost $\tilde L^\gamma$ in the undiscounted setting with prediction horizon $1$ and terminal cost $\tilde V_{\vect{\pi}_\star^\gamma}^\gamma$. Using~\eqref{eq:Vtilde}, we obtain that $\tilde V_{\vect{\pi}_\star^\gamma}^\gamma$ is the resulting value function, such that, by propagating the reasoning over several steps, by induction we obtain
	\begin{equation*}
		\vect{\pi}_\star^\gamma \left (\vect{x}_0\right ) = \arg\min_{\vect{\pi}\in\Pi} \, \lim_{N\to\infty} \sum_{k=0}^{N-1} \tilde L^\gamma \left (\vect{x}_k^\vect{\pi},\vect{\pi}\left (\vect{x}_k^\vect{\pi}\right )\right ) + \tilde V_{\vect{\pi}_\star^\gamma}^\gamma \left ( \vect{x}_N^\vect{\pi}\right ).
	\end{equation*}
	Note that $\tilde V_{\vect{\pi}_\star^\gamma}^\gamma(\vect{x})=V_\star^\gamma(\vect{x})$ is continuous at $\vect{x}_\mathrm{s}$ by Assumption~\ref{ass:controllability} and $\tilde V_{\vect{\pi}_\star^\gamma}^\gamma(\vect{x}_\mathrm{s})=0$. Moreover,  by Theorem~\ref{thm:asymptotic_stability} policy $\vect{\pi}_\star^\gamma$ yields asymptotic stability. Therefore, we also have
	\begin{subequations}
		\label{eq:undisc_prob_constr}
		\begin{align}
			\vect{\pi}_\star^\gamma \left (\vect{x}_0\right ) = \arg\min_{\vect{\pi}\in\Pi} \ &\sum_{k=0}^\infty \tilde L^\gamma \left (\vect{x}_k^\vect{\pi},\vect{\pi}\left (\vect{x}_k^\vect{\pi}\right )\right ) \\
			\mathrm{s.t.} \ & \lim_{k\to\infty} \vect{\vect{x}_k^\vect{\pi}} = \vect{x}_\mathrm{s}.
		\end{align}
	\end{subequations}
	We now observe that, by Assumption~\ref{ass:disc_str_diss}, Equation~\eqref{eq:discounted_dissipativity2} entails that the undiscounted strict dissipativity criterion \eqref{eq:strict_diss} holds for the undiscounted problem~\eqref{eq:pitilde}. This directly entails that $\vect{\tilde \pi}_\star^\gamma$, solution of~\eqref{eq:pitilde}, must asymptotically stabilize the system to the steady state $\vect{x}^\mathrm{s}$. Consequently, $\vect{\tilde \pi}_\star^\gamma$ is a feasible solution of~\eqref{eq:undisc_prob_constr}, which entails that, by optimality of $\vect{\pi}_\star^\gamma$ for~\eqref{eq:undisc_prob_constr}, we must have $\tilde V_{\vect{\tilde \pi}_\star^\gamma}^\gamma(\vect{x}) \geq \tilde V_{\vect{\pi}_\star^\gamma}^\gamma(\vect{x})$. This proves that $\vect{\pi}_\star^\gamma(\vect{x})$ must also be optimal for~\eqref{eq:pitilde} 
	and, consequently,
	\begin{align*}
		\tilde V_\star^\gamma (\vect{x}) =\tilde V_{\vect{\pi}_\star^\gamma}^\gamma (\vect{x}) \overset{\eqref{eq:Vhat_V}}{=} V_\star^\gamma (\vect{x}).
	\end{align*}
	$\hfill\qed$
\end{pf}

By exploiting this equivalence, Condition~\eqref{eq:discounted_dissipativity2} can be interpreted as the standard strict dissipativity condition on the rotated stage cost~\cite{Amrit2011a}, i.e.,:
\begin{align}
	\tilde L^\gamma(\vect{x},\vect{u}) + \lambda(\vect{x}) - \lambda(\vect{f}(\vect{x},\vect{u})) \geq \rho(\|\vect{x}-\vect{x}^\mathrm{s}\|).
\end{align}
Note, however, that condition~\eqref{eq:discounted_dissipativity2} alone is not sufficient for asymptotic stability in the discounted case, since condition~\eqref{eq:discounted_dissipativity1} is necessary in order to establish the equivalence proven in Theorem~\ref{thm:undiscounted_equivalence} and the lower bound for $\hat V_\star^\gamma$. 

\subsection{Discounted Optimal Steady-State}

We define next the optimal steady state to which the closed-loop system converges, provided that the SDSD Assumption~\ref{ass:disc_str_diss} is met. 
To that end, we use stage cost $\tilde L^\gamma$ to formulate the optimization problem
\begin{subequations}%
	\label{eq:discounted_optimal_steady_state}
	\begin{align}
	(\vect{x}^\mathrm{s}_\star,\vect{u}^\mathrm{s}_\star) := \arg \min_{\vect{x},\vect{u}} \  & 
	\tilde L^\gamma (\vect{x},\vect{u}) \\
	\mathrm{s.t.} \ & \vect{x} = \vect{f}(\vect{x},\vect{u}).
	\end{align}
\end{subequations}
Note that, in general, $(\vect{x}^\mathrm{s}_\star,\vect{u}^\mathrm{s}_\star)$ needs not be unique.
We prove in the next theorem that, if the SDSD Assumption~\ref{ass:disc_str_diss} holds for some steady state, then it must hold for steady-state~\eqref{eq:discounted_optimal_steady_state}. 

\begin{theorem}
	\label{thm:steady_state}
	Suppose that Assumptions~\ref{ass:regularity}, \ref{ass:disc_str_diss}, \ref{ass:controllability} hold for a given $(\vect{x}^\mathrm{s},\vect{u}^\mathrm{s})$. Then, $(\vect{x}^\mathrm{s},\vect{u}^\mathrm{s})$ must be a solution of~\eqref{eq:discounted_optimal_steady_state}:
	\begin{align}
	(\vect{x}^\mathrm{s},\vect{u}^\mathrm{s}) \in (\vect{x}^\mathrm{s}_\star,\vect{u}^\mathrm{s}_\star).
	\end{align}
\end{theorem}
\begin{pf}
	By Theorem~\ref{thm:asymptotic_stability}, we have that, from any initial state, the system in closed-loop with policy $\vect{\hat \pi}_\star^\gamma (\vect{x})=\vect{\pi}_\star^\gamma (\vect{x})$ is stabilized to $\vect{x}^\mathrm{s}=\vect{f}(\vect{x}^\mathrm{s},\vect{u}^\mathrm{s})$. Assume by contradiction that $(\vect{x}^\mathrm{s},\vect{u}^\mathrm{s})\notin(\vect{x}^\mathrm{s}_\star,\vect{u}^\mathrm{s}_\star)$. By definition of $(\vect{x}^\mathrm{s}_\star,\vect{u}^\mathrm{s}_\star)$, this entails that 
	\begin{align}
	\label{eq:absurd_result}
	0=\tilde L^\gamma (\vect{x}^\mathrm{s},\vect{u}^\mathrm{s})>\tilde L^\gamma (\vect{\bar x},\vect{\bar u}),  && \forall \ (\vect{\bar x},\vect{\bar u})\in (\vect{x}^\mathrm{s}_\star,\vect{u}^\mathrm{s}_\star).
	\end{align}
		The trajectory $\vect{x}_k=\vect{\bar x}$, $\vect{u}_k=\vect{\bar u}$, $k=0,\ldots,\infty$ is feasible for Problem~\eqref{eq:pitilde} and, by~\eqref{eq:absurd_result}, yields the cost
		\begin{align*}
			\sum_{k=0}^{\infty} \tilde L^\gamma(\vect{\bar x},\vect{\bar u}) = -\infty,
	\end{align*}
	such that $\tilde V_{\vect{\pi}_\star^\gamma}^\gamma (\vect{\bar x})=-\infty$.
	
	However, because $\left |\tilde L^\gamma (\vect{\bar x},\vect{\bar u})\right |<\infty$, we have that $\vect{\bar x}\in\mathbb{X}_0$. By Theorem~\ref{thm:undiscounted_equivalence} we have $\tilde V_{\vect{\pi}_\star^\gamma}^\gamma (\vect{\bar x})=V_\star^\gamma (\vect{\bar x})$ and Assumption~\ref{ass:controllability} guarantees that $V_\star^\gamma (\vect{x})$ is bounded for all $\vect{x}\in\mathbb{X}_0$, which yields the desired contradiction.
	$\hfill\qed$
\end{pf}
This theorem entails that in the general case the optimal steady-state depends on the discount factor $\gamma$, i.e., since by~\eqref{eq:undiscounted_equivalent_stage_cost} $\tilde L^\gamma$  does depend on $\gamma$, different $\gamma$ can yield different $(\vect{x}^\mathrm{s}_\star,\vect{u}^\mathrm{s}_\star)$, as had been conjectured in~\cite{Gaitsgory2018} and discussed in~\cite{Gruene2020}.

\begin{Remark}
	We observe that, if~\eqref{eq:discounted_dissipativity} holds for $(\vect{x}^\mathrm{s},\vect{u}^\mathrm{s})$, then $\tilde L^\gamma(\vect{x},\vect{u}) \geq \rho\left (\left \| \vect{x}-\vect{x}^\mathrm{s} \right \|\right )$, such that $\vect{x}^\mathrm{s}_\star$ must be unique. However, without stronger assumptions, e.g., using $\rho(\|\vect{x}-\vect{x}^\mathrm{s}, \ \vect{u}-\vect{u}^\mathrm{s} \|)$ in~\eqref{eq:discounted_dissipativity}, the optimal input $\vect{u}^\mathrm{s}_\star$ need not be unique. 
\end{Remark}

Note that the result of Theorem~\ref{thm:steady_state} is not surprising in the light of the equivalence with the undiscounted setting stated in Theorem~\ref{thm:undiscounted_equivalence}. Since undiscounted MPC with stage cost $\tilde L^\gamma$ is equivalent to discounted MPC with stage cost $L$, the steady-state they converge to must also match. Indeed,~\eqref{eq:discounted_optimal_steady_state} yields the optimal steady state for the standard case of undiscounted economic MPC, such that the proof of Theorem~\ref{thm:steady_state} can equivalently be obtained by combining Theorem~\ref{thm:undiscounted_equivalence} with the standard results for the undiscounted case.

After having analyzed the properties that stem from SDSD, we comment next on the relationship between SDSD and other conditions used in the literature.

\subsection{Relationship with Other Conditions for Stability}
\label{sec:stability_conditions}

Since the setting of this paper has been analyzed in~\cite{Gaitsgory2018} using different assumptions, we discuss next the main differences between our assumptions and the ones therein. In both cases it is assumed that~\eqref{eq:discounted_dissipativity1} holds, i.e.,
\begin{align*}
	\hat L^\gamma(\vect{x},\vect{u}) \geq \rho\left ( \left \| \vect{x}-\vect{x}^\mathrm{s} \right \|\right ).
\end{align*}
The main difference is that we assume~\eqref{eq:discounted_dissipativity2}, while in~\cite{Gaitsgory2018} it is required that, given two constants $0\leq \varphi<\Phi$, for all 
\begin{align*}
	\varphi \leq \| \vect{x} - \vect{x}^\mathrm{s} \| \leq \Phi,
\end{align*}
there exists another constant $1\leq C<(1-\gamma)^{-1}$ such that
\begin{align}
	\label{eq:conservative_condition}
	\hat V_\star^\gamma (\vect{x}) \leq C \inf_{\vect{u}} \hat L^\gamma(\vect{x},\vect{u}).
\end{align}
An alternative less conservative condition proposed in~\cite{Gaitsgory2018} is
\begin{align}
	\label{eq:conservative_condition1}
	\hat V_\star^\gamma (\vect{x}) \leq C \hat L^\gamma(\vect{x},\vect{\pi}_\star^\gamma(\vect{x})) ,
\end{align}
with $1\leq C<(1-\gamma)^{-1}$. In order to be able to compare~\eqref{eq:discounted_dissipativity2} with~\eqref{eq:conservative_condition1}, we first rewrite both in equivalent expressions which allow us to better highlight the similarities and differences.
By selecting $\varphi>0$ arbitrarily small, the strict inequality holds with equality only for $\vect{x}=\vect{x}_\mathrm{s}$. By further selecting the most favorable constant $C=(1-\gamma)^{-1}$ one can replace~\eqref{eq:conservative_condition1} with
\begin{align}
\label{eq:conservative_condition2}
	&(1-\gamma)\hat V_\star^\gamma (\vect{x}) \leq \hat L^\gamma(\vect{x},\vect{\pi}_\star^\gamma(\vect{x})) -  \rho^\prime(\|\vect{x}-\vect{x}^\mathrm{s}\|),
\end{align}
with $\rho^\prime\in\mathcal{K}$. 
Note that we exploited the fact that $\hat L^\gamma(\vect{x},\vect{\pi}(\vect{x})) > 0$ for all $\vect{x}\neq\vect{x}_\mathrm{s}$, while in~\cite{Gaitsgory2018} the inequality was applied to the original value function $V_\star^\gamma$ instead of the modified $\hat V_\star^\gamma$. Finally, we observe that, using the Bellman equation, i.e., Equation~\eqref{eq:bellman} written for stage cost $\hat L^\gamma$, in~\eqref{eq:conservative_condition2} we have
\begin{align*}
	&(1-\gamma)\left (\hat L^\gamma(\vect{x},\vect{\pi}_\star^\gamma(\vect{x})) + \gamma \hat V_\star^\gamma (\vect{f}(\vect{x}, \vect{\pi}_\star^\gamma(\vect{x})))\right ) \\
	&\hspace{10em}\leq \hat L^\gamma(\vect{x},\vect{\pi}_\star^\gamma(\vect{x})) -  \rho^\prime(\|\vect{x}-\vect{x}^\mathrm{s}\|),
\end{align*}
which can be rewritten as
\begin{align}
	\label{eq:conservative_condition3}
	&(1-\gamma)\hat V_\star^\gamma (\vect{f}(\vect{x}, \vect{\pi}_\star^\gamma(\vect{x}))) \leq  \hat L^\gamma(\vect{x},\vect{\pi}_\star^\gamma(\vect{x})) -  \rho(\|\vect{x}-\vect{x}^\mathrm{s}\|),
\end{align}
where $\gamma\rho(\cdot)= \rho^\prime(\cdot)$.

Condition~\eqref{eq:discounted_dissipativity2}, using~\eqref{eq:Lhat} and~\eqref{eq:V_Vhat}, equivalently reads
\begin{align}
	\label{eq:discounted_dissipativity2_twisted}
	(1-\gamma)\hat V_\star^\gamma (\vect{f}(\vect{x}, \vect{u})) \leq \hat L^\gamma(\vect{x},\vect{u}) -  \rho(\|\vect{x}-\vect{x}^\mathrm{s}\|).
\end{align}
We can now compare our assumption with that of~\cite{Gaitsgory2018}, since~\eqref{eq:conservative_condition3} closely resembles~\eqref{eq:discounted_dissipativity2_twisted}, except for the fact that the functions in~\eqref{eq:conservative_condition3} are evaluated using the optimal policy. Indeed, in Equation~\eqref{eq:lyap_decrease} in the proof of Theorem~\ref{thm:asymptotic_stability} we only need that~\eqref{eq:conservative_condition3} holds for the optimal policy and not for all possible control inputs. Consequently, Assumption~\ref{ass:disc_str_diss} could be relaxed to only hold for $\vect{u}=\vect{\pi}_\star^\gamma(\vect{x})$, therefore obtaining a condition which closely resembles the one used in~\cite{Gaitsgory2018}.

A similar situation is found in the undiscounted case, where, though the stability proof only requires strict dissipativity to hold for the optimal policy, it is commonly assumed that strict dissipativity holds for all feasible inputs. However, this is not restrictive: it has been proven in~\cite{Mueller2015a} that, under a mild controllability assumption, if $(\vect{x}^\mathrm{s},\vect{u}^\mathrm{s})\in\mathrm{int}\mathbb{Z}$, strict dissipativity is also necessary for asymptotic stability.  Investigating whether this also applies to the discounted case has been partially done in~\cite{Mueller2017} and will be the subject of future research. 
In this paper, we limit ourselves to the next discussion, which exploits the equivalence with the undiscounted case. 

In order to provide further insight about Assumption~\eqref{eq:discounted_dissipativity2}, we state the following lemma.

\begin{Lemma}
	\label{lem:undiscounted_equivalence}
	Suppose that Assumption~\ref{ass:regularity} and
	\begin{align}
		\label{eq:weak_stability_assumption}
		\lim_{N\to\infty} V_\star^\gamma\left (\vect{x}_N^{\vect{\pi}_\star^\gamma}\right ) =0, && \left |V_\star^\gamma\left (\vect{x}_k^{\vect{\pi}_\star^\gamma}\right )\right | < \infty, \quad \forall \ k \geq 0
	\end{align}
	hold. Then, 
	\begin{align}
	\vect{\tilde \pi}_\star^\gamma(\vect{x}) = \vect{ \pi}_\star^\gamma(\vect{x}), &&  \tilde V_\star^\gamma(\vect{x})= V_\star^\gamma(\vect{x}).
	\end{align}
\end{Lemma}
\begin{pf}
	The proof follows the same arguments used in Theorem~\ref{thm:undiscounted_equivalence}. $\hfill\qed$
\end{pf}

We observe that, if~\eqref{eq:weak_stability_assumption} does not hold, then the closed-loop system cannot be asymptotically stabilized to $(\vect{x}^\mathrm{s},\vect{u}^\mathrm{s})$ by policy $\vect{\pi}_\star^\gamma$, such that this assumption is not conservative in this context. Note however, that~\eqref{eq:weak_stability_assumption} is not sufficient for asymptotic stability, since we do not assume that $V_\star^\gamma$ neither its evolution in time is upper or lower bounded by class $\mathcal{K}$ functions.

By Lemma~\ref{lem:undiscounted_equivalence}, there exists an undiscounted formulation which yields the same optimal policy and value function as the discounted formulation. This equivalent formulation can be obtained by modifying stage cost $\hat L^\gamma$ and, consequently, the corresponding value function $\hat V_\star^\gamma$, 
to obtain the undiscounted stage cost
\begin{align*}
	\hat{\tilde L}^\gamma(\vect{x},\vect{u}) :=  \hat L^\gamma (\vect{x},\vect{u}) + (\gamma -1)\hat V_\star^\gamma (\vect{f}(\vect{x}, \vect{u})),
\end{align*}
yielding value function $\hat{\tilde{V}}_\star^\gamma$, obtained by replacing ${\tilde L}^\gamma$ with $\hat{\tilde L}^\gamma$ in~\eqref{eq:pitilde}-\eqref{eq:Vtilde}. 
Note that in the literature on undiscounted economic MPC, $\hat{\tilde L}^\gamma$ is called a rotated cost (with respect to $\tilde L^\gamma$) and used  to prove asymptotic stability~\cite{Amrit2011a}, similarly to what we did using $\hat L^\gamma$ in the discounted setting.
We exploit this equivalence to discuss next possible relaxations of Assumption~\ref{ass:disc_str_diss} in the undiscounted setting.

If both~\eqref{eq:weak_stability_assumption} and Assumptions~\ref{ass:regularity},~\ref{ass:disc_str_diss}(i),~\ref{ass:controllability} hold, then the equivalent undiscounted formulation yields a value function which is both upper and lower bounded by class $\mathcal{K}$ functions, but the stage cost $\hat{\tilde L}^\gamma$ is not guaranteed to be positive definite, such that we cannot directly prove the decrease condition for the Lyapunov function candidate $\hat{\tilde{V}}_\star^\gamma$. The missing condition, i.e., positive-definiteness of the stage cost is precisely Assumption~\ref{ass:disc_str_diss}(ii).

On the other hand, if both~\eqref{eq:weak_stability_assumption} and Assumptions~\ref{ass:regularity},~\ref{ass:disc_str_diss}(ii),~\ref{ass:controllability} hold, then the stage cost  $\hat{\tilde L}^\gamma$ is positive definite, which, in turn, implies the Lyapunov lower bound on the value function $\hat{\tilde{V}}_\star^\gamma$. Since the Lyapunov upper bound is granted by Assumption~\ref{ass:controllability}, in this case we do have that $\hat{\tilde{V}}_\star^\gamma$ is a Lyapunov function. Since~\eqref{eq:weak_stability_assumption} and Assumption~\ref{ass:regularity} guarantee that the discounted and undiscounted formulations are equivalent and the remaining assumptions are standard in the undiscounted setting, this is possibly the least restrictive set of assumptions to prove asymptotic stability in the discounted setting. However, as discussed above,~\eqref{eq:weak_stability_assumption} is already a weak stability assumption in itself. 

\section{Examples}
\label{sec:examples}

We provide next two examples which we intentionally select as simple enough in order to be able to provide insight in  the theoretical results of this paper. 

\subsection{Linear Quadratic Regulator (LQR)}
Consider the simple example from~\cite{Gaitsgory2018}, with linear dynamics
\begin{align*}
	\vect{x}_+ &= A\vect{x} + B\vect{u},\\
	A &= \matr{cc}{2 & 0 \\ 1 & 2}, & B&=\matr{cc}{1 & 0 \\ 0 & 1},
\end{align*}
and quadratic stage cost
\begin{align*}
	L(\vect{x},\vect{u}) &= \vect{x}^\top Q \vect{x} + \vect{u}^\top R \vect{u},\\
	Q&=\matr{cc}{1 & 0 \\ 0 & 1}, & R&=\matr{cc}{1 & 0 \\ 0 & 1}.
\end{align*}
In order to abide by the proposed theory, state and control constraints ought to be added. We observe here that for any choice of such constraints which defines a sufficiently large region having the origin in its interior, there will be a set of initial conditions such that the constraints are never active over the LQR trajectories. We additionally observe that this set of initial conditions can be made arbitrarily large by increasing the size of the feasible state and input set defined by these constraints. It follows that the proposed theory applies within an arbitrarily large set of initial conditions, including the origin. Due to these observations, we do not specify state and input constraints here.
Note however that, in case such constraints are removed, it is possible that the optimal solution is not stabilizing even though SDSD holds.

In~\cite{Gaitsgory2018} it has been noted that a constant $C$ satisfying~\eqref{eq:conservative_condition} with $\lambda(\vect{x})=0$ exists only for $\gamma$ larger than $\approx0.846$, but $V_\star^\gamma(\vect{x})=\vect{x}^\top P \vect{x}$ is a Lyapunov function for all $\gamma$ larger than $\approx0.3342$. Nevertheless, the system is still stable for $\gamma=0.334$, even though $V_\star^\gamma$ is not a Lyapunov function. 

In this example, since the dynamics are linear and the cost is quadratic, we are able to compute a quadratic function $\lambda(\vect{x})=\vect{x}^\top \Lambda \vect{x}$ satisfying conditions~\eqref{eq:discounted_dissipativity} by solving a Semidefinite Program (SDP), which we solve using Yalmip~\cite{Lofberg2004} and Mosek~\cite{MOSEK-hp}. 
For $\gamma=0.334$, this yields (up to the fifth significant digit),
\begin{align*}
	\Lambda =-\matr{cc}{3.9511 & 1.2702 \\ 1.2702 & 2.7738},
\end{align*}
with corresponding modified stage cost
\begin{align*}
	&\hat L^\gamma(\vect{x},\vect{u}) = \matr{c}{\vect{x}\\ \vect{u}}^\top \hat H^\gamma \matr{c}{\vect{x}\\ \vect{u}}, \\
	&\hat H^\gamma = \matr{cccc}{4.3127 &  1.9906 &  2.8167 &  1.6304 \\
		1.9906   & 1.6905  &  0.7643   & 1.7322 \\
		2.8167  &  0.7643   & 2.2173   & 0.3822 \\
		1.6304   & 1.7322   & 0.3822  &  1.8661 } \succ 0.023 I.\nonumber 
\end{align*}
Then, we obtain a Lyapunov function as
\begin{align*}
	\hat V_\star^\gamma(\vect{x}) = V_\star^\gamma(\vect{x}) + \lambda(\vect{x}) = \vect{x}^\top \hat P^\gamma \vect{x},
\end{align*}
with
\begin{align*}
	\hat P^\gamma = P + \Lambda = \matr{cc}{0.1537 & 0.0632 \\ 0.0632 & 0.0907} \succ 0.0516 I,
\end{align*}
and $P$ the matrix defining $V_\star^\gamma(\vect{x})=\vect{x}^\top P \vect{x}$.

Finally, the solution is stabilizing for $\gamma\geq \underline{\gamma}$, with $\underline{\gamma}\approx0.3109$. For ${\gamma}=0.3109$ we can still compute a matrix $\Lambda$ which satisfies SDSD and yields
\begin{align*}
	\hat H^\gamma \succ 1.325 \cdot 10^{-8} I, && \hat P^\gamma \succ 2.459 \cdot 10^{-4} I,
\end{align*}
such that our conditions are tight for this example. 

Note that, since $H\succ0$, condition~\eqref{eq:discounted_dissipativity1} is satisfied for $\Lambda=0$, but this is not enough to conclude asymptotic stability. Indeed, condition~\eqref{eq:discounted_dissipativity2} is not satisfied for $\Lambda=0$ for $\gamma \leq \gamma_1$ with $\gamma_1\approx 0.93507$, even though, as pointed out above, $V_\star^\gamma$ is a Lyapunov function for $\gamma$ larger than $\approx0.3342$. 

Finally, we stress that for $\gamma<\underline{\gamma}\approx0.3109$, if we impose either only~\eqref{eq:discounted_dissipativity1} or only~\eqref{eq:discounted_dissipativity2}, we are always able to compute a storage function, even though the closed-loop system is not asymptotically stable. This fact highlights the well-known insufficiency of discounted strict dissipativity for proving stability, unless further conditions are met.

\subsection{A Simple Nonlinear Example}

Consider the scalar nonlinear system defined by
\begin{align*}
	\vect{x}_+ &= 0.01\,\vect{u}(1-\vect{x}) + 0.96\,\vect{x}, \\
	L(\vect{x},\vect{u}) &= -1.5\,\vect{u} + 2\,\vect{u}\vect{x} + 0.1\,(\vect{u}-4)^2,
\end{align*}
with $\vect{x}\in [0,1] \subset \mathbb{R}$,  $\vect{u}\in [0,20] \subset \mathbb{R}$.

Since the problem is of sufficiently small dimension, we can solve it by dynamic programming. We verified numerically that the storage function
\begin{align*}
	\lambda(\vect{x}) = -\nabla_\vect{x} V_\star^\gamma(\vect{x}^\mathrm{s}) \, (\vect{x}-\vect{x}^\mathrm{s}) + 50 \, (\vect{x}-\vect{x}^\mathrm{s})^2,
\end{align*}
satisfies the SDSD assumption for all discount factors $\gamma\in \ ]0,1]$. 
Indeed, in this case the policy $\vect{\pi}_\star^\gamma$ is stabilizing for all discount factors $\gamma$ in that interval. 

\begin{figure}
	\includegraphics[width=\linewidth]{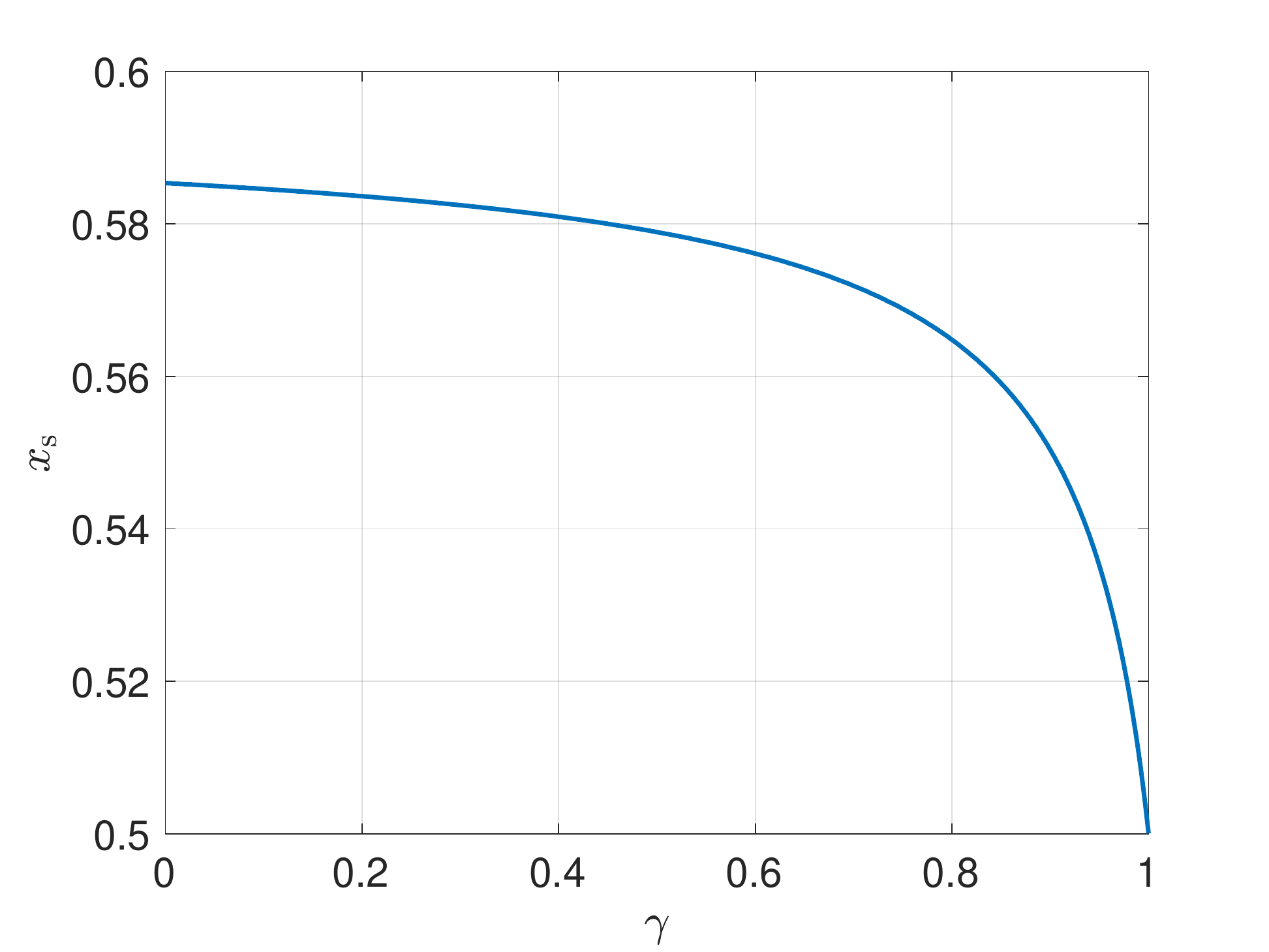}
	\caption{Optimal steady state as a function of $\gamma$.}
	\label{fig:cstr_ss}
\end{figure}

We display in Figure~\ref{fig:cstr_ss} the optimal steady state as a function of the discount factor $\gamma$, where we evaluated numerically that the optimal policy indeed stabilizes the system to the solution of the steady-state problem~\eqref{eq:discounted_optimal_steady_state}, which we solved numerically.

\section{Conclusions}
\label{sec:conclusions}

In this paper we have defined a new concept of dissipativity for discounted MPC, which allowed us to prove asymptotic stability, to characterize the optimal steady-state and to establish an equivalence between discounted and undiscounted problems. We have compared our assumptions to similar ones used in the literature and we have provided two examples to illustrate our findings.

Future work will consider further investigating the necessity of our assumption. Furthermore, the finite-horizon case presents some challenges which impede a direct application of the available stability theory. Similarly, the stochastic case, which is particularly relevant in the context of reinforcement learning, is far from being fully understood.

\bibliographystyle{plain}
\bibliography{syscop}

\end{document}